\documentclass[aps,twocolumn,pra,showpacs,floatfix]{revtex4}

\usepackage{epsfig}
\usepackage{graphicx}
\usepackage{dcolumn}
\usepackage{amsmath}

\begin{document}

\title{The effect of atomic electrons on nuclear fission}

\author{V. A. Dzuba and V. V. Flambaum}

\affiliation{School of Physics, University of New South Wales,
Sydney 2052, Australia}

\date{\today}

\begin{abstract}
We calculate correction to the nuclear fission barrier produced by the atomic
electrons. The result presented in analytical form is convenient to use in
future nuclear calculations. The atomic electrons have a small stabilizing
effect on nuclei, increasing lifetime in nuclear fission channel. This effect
gives a new instrument to study the fission process.
\end{abstract}

\pacs{24.75.+i,31.30.Gs}

\maketitle

 In this paper we consider the effect of atomic
electrons on nuclear fission. This effect is small for all known nuclei.
However, it might still be detectable. It can be observed by comparing
fission probabilities of nuclei striped from all electrons with those
in neutral atoms. This may reveal important information about fission.
The effect is expected to play more significant role in fission of 
superheavy nuclei leading to increased lifetime of the nuclei.
The study of the superheavy elements is now a popular area of research due 
to the search for the hypothetical {\em stability island} ($Z$=114 to 126)
and good progress in the synthesis of the elements on accelerators up to
$Z=118$ (see, e.g. reviews~\cite{Greiner,Oganessian,Cwiok,Ackermann}).
It is natural to expect that physics
of the superheavy nuclei has some phenomena which never manifest themselves
for lighter nuclei.

In this paper we perform relativistic atomic many-body calculations and give a
formula which fits calculated total energy of an atom as a function
of nuclear charge $Z$ and nuclear radius $R_N$. This formula can
be used in  nuclear calculations of the fission probabilities.

Nuclear fission is the main decay channel for heavy nuclei starting from 
$^{254}_{98}$Cf~\cite{Segre}. During the
fission a nucleus goes throw several stages of deformation before separation
of the fragments. The process is accompanied by the nuclear rotation.

 The electrons are far away from the nucleus and the electron density does not
rotate when the nucleus rotate. Thus, the electrons feel nuclear charge density
averaged over the nuclear rotation. This nuclear charge density for spinless
nucleus is spherical.
For a non-zero nuclear spin there is an electric quadrupole moment correction
which is negligible anyway. Here there is the difference with the muon case
considered in Ref. \cite{Leander}: the muon is inside the nucleus, so the
 calculation in the rotating (frozen nucleus) frame is appropriate.

Therefore, the change of the electron energy during the fission process
is similar to the well known volume (or field) isotope shift of atomic
 transition frequencies which is determined by a change
of a single nuclear parameter, mean squared nuclear charge radius.
Total electron energy is negative and it goes up when nuclear radius is
increasing. This means
that electrons take some energy from the process making fission more
difficult.
\begin{figure}
\centering
\epsfig{figure=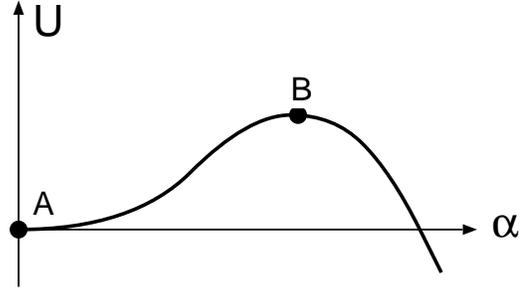,scale=1.2}
\caption{Nuclear energy $U(\alpha)$ as a function of deformation parameter $\alpha$.}
\label{fig}
\end{figure}

Stress that in this paper we do not perform any nuclear calculations.
A real shape of the fission barrier is complicated, with several minima, and
an accurate calculation would require sophisticated nuclear codes.
However, to explain the physics involved and find out if the measurements
 are feasible we start from presenting some rough estimates using
 a simplest  parabolic barrier
 model (Fig.~\ref{fig})~\cite{Segre,Bohr}. The probability of the tunneling
through the barrier is given by~\cite{Bohr}
\begin{equation}
  P = \frac{1}{1+\exp\left(-2\pi\frac{E-U_B}{\hbar \omega}\right)},
\label{P}
\end{equation}
where $U_B$ is the maximum of the potential  energy,
 (point B at  Fig.~\ref{fig}) and $\hbar \omega$
is of the order from 0.5 to 1 MeV~\cite{Bohr}. For the spontaneous fission
the tunneling amplitude is very small. Indeed, according to ref.~\cite{Segre} 
$U_B-U_A \sim 5 \rm{MeV}$. Therefore, $2\pi|U_A-U_B|/\hbar \omega \gg 1$
and
\begin{equation}
  P \approx \exp\left(2\pi\frac{E-U_B}{\hbar \omega}\right) \equiv P_0
\label{P2}
\end{equation}
for energies not too close to $U_B$.
 To estimate the effect
of electrons on nuclear fission we should take into account
the  electron energy $E_e(\alpha)=E_e(\alpha_A)+k (\alpha-\alpha_A) + ...$
 where $\alpha$ is the deformation parameter
(see  Fig.~\ref{fig}). The constant $E_e(\alpha_A)$ does not change
the probability of the tunneling since the constant potential
does not influence the wave function  (the bound state energy $E$ and the
 potential $U$ change by the same amount $E_e(\alpha_A)$ so the difference
 $E-U_B$ does not change).
The linear term $k (\alpha-\alpha_A)$
can be incorporated into  the effective  parabolic potential and
 slightly changes the difference $E-U_B$. Indeed, it vanishes at the point
 of minimum $ \alpha=\alpha_A$ where the bound state is located, and modify
the potential near the maximum which determines the tunneling probability.
A simple calculation gives the following result for the tunneling probability  
\begin{equation}
  P = P_0 \exp\left(-2\pi\frac{\delta E}{\hbar \omega}\right).
\label{Pe}
\end{equation}
where  $\delta E=E_e(\alpha_B)-E_e(\alpha_A)$ is the change
of the electron energy.
 This change is positive and therefore
the factor in (\ref{Pe}) is smaller than one. This means that the electrons
make the probability of nuclear fission smaller.

\begin{table*}
\caption{Energy of $1s$ electron and total electron energy (MeV) of many-electron atoms as a function
of nuclear charge $Z$ and nuclear radius $r_N$. In the last column we present
rough estimates of the relative change of fission probability assuming
 10\% change of nuclear radius between points A and B; the actual increase with
nuclear charge is slower since the relative change of radius in heavy
 elements is smaller than in light elements.}
\label{tb:e}
\begin{ruledtabular}
\begin{tabular}{r|rrr|rrr|rr|c}
\multicolumn{1}{c|}{$Z$} &
\multicolumn{1}{c}{$r_N$} &\multicolumn{2}{c|}{Energy [MeV]} &
\multicolumn{1}{c}{$r_N$} &\multicolumn{2}{c|}{Energy [MeV]} &
\multicolumn{1}{c}{$\Delta_{1s}$} &\multicolumn{1}{c|}{$\Delta_{tot}$} &
\multicolumn{1}{c}{$|P-P_0|/P_0$} \\
&\multicolumn{1}{c}{[fm]} &\multicolumn{1}{c}{1s} & \multicolumn{1}{c|}{Total} & 
\multicolumn{1}{c}{[fm]} &\multicolumn{1}{c}{1s} &\multicolumn{1}{c|}{Total} & 
\multicolumn{1}{c}{[MeV]} &\multicolumn{1}{c|}{[MeV]} &
\multicolumn{1}{c}{[\%]} \\
\hline
 80 &  6.44 & -0.083656 & -0.534682 &  7.09 & -0.083649 & -0.534664 &  0.00001 &  0.00002 & 0.02 \\
 90 &  7.75 & -0.110411 & -0.721344 &  8.53 & -0.110387 & -0.721283 &  0.00002 &  0.00006 & 0.08 \\
100 &  8.03 & -0.142969 & -0.948529 &  8.84 & -0.142908 & -0.948367 &  0.00006 &  0.00016 & 0.20 \\
110 &  8.29 & -0.183076 & -1.225050 &  9.12 & -0.182917 & -1.224613 &  0.00016 &  0.00044 & 0.55 \\
120 &  8.54 & -0.233429 & -1.563897 &  9.39 & -0.233024 & -1.562712 &  0.00041 &  0.00118 & 1.5  \\
130 &  8.76 & -0.298439 & -1.987579 &  9.64 & -0.297409 & -1.984307 &  0.00103 &  0.00327 & 4.0  \\
140 &  8.99 & -0.385348 & -2.538006 &  9.89 & -0.382769 & -2.528802 &  0.00258 &  0.00920 & 11   \\
150 &  9.20 & -0.505625 & -3.289468 & 10.12 & -0.499261 & -3.263392 &  0.00636 &  0.02608 & 28   \\
160 &  9.40 & -0.675140 & -4.345553 & 10.34 & -0.660488 & -4.278862 &  0.01465 &  0.06669 & 57   \\
\end{tabular}
\end{ruledtabular}
\end{table*}


To calculate $\delta E$ we need to calculate total electron energy of an 
atom for two different nuclear charge distributions corresponding to
point A and point B  (Fig.~\ref{fig}). In light actinide atoms,
 e.g. $^{236}U$, the deformation changes
from 0.2 in the  minimum to about 1 in the last maximum of the
 fission barrier (see e.g. \cite{Segre,Leander}).
In heavier elements the difference is
significantly smaller, about 0.2-0.4. As we mentioned above,
after the averaging over nuclear rotation the charge distribution
is spherically symmetric.
For an estimate we assume that the change of the radius of the nuclear charge
distribution from point A to point B is 10\%
 (an accurate calculation will be discussed below). 
We calculate nuclear potential by integrating standard Fermi distribution
for nuclear density
\begin{equation}
  \rho(r) = \frac{C}{1+exp\left(\frac{r-r_N}{D}\right)},
\label{rho}
\end{equation}
where $D=d/4\ln3$, $d$=2.3 fm, $r_N \approx 1.2(3Z)^{1/3}$ fm and $C$ is
a normalization factor defined by $\int \rho dV =Z|e|$.

To find the electron energy of the atom we solve self-consistently a
set of relativistic Hartree-Fock equations for single-electron orbitals
(atomic units)
\begin{equation}
    \begin {array}{c} \dfrac{df_i}{dr}+\dfrac{\kappa_{i}}{r}f_v(r)-
    \left[2+\alpha^{2}(\epsilon_{i}-\hat{V})\right]g_i(r)=0,  \\[0.5ex]
    \dfrac{dg_a}{dr}-\dfrac{\kappa_{i}}{r}f_i(r)+(\epsilon_{i}-
    \hat{V})f_i(r)=0, \end{array}
\label{Dirac}
\end{equation}
here $\kappa=(-1)^{l+j+1/2}(j+1/2)$,$l$ and $j$ are angular and total
electron momenta
and $\hat{V}$ is the sum of nuclear potential, found by integration of 
nuclear density (\ref{rho}) and the self-consistent Hartree-Fock potential 
of atomic electrons. Index $i$ numerates single-electron states.

Total electron energy in first-order in Coulomb interaction
is given by
\begin{equation}
  E_{total} = \sum_{i=1}^{Z} \epsilon_i - \sum_{i<j} \tilde{q}(ijij),
\label{total}
\end{equation}
where $\epsilon_i$ are Hartree-Fock energies (Eq.~(\ref{Dirac})) 
of $Z$ atomic electrons,
$\tilde{q}(ijij) = q(ijij)-q(ijji)$ and $q(abcd)$ is a Coulomb integral
\[ q(abcd) = \int\int \psi_a(r_1)\psi_b(r_2)\frac{e^2}{|\mathbf{r}_1-
\mathbf{r}_2|}\psi_c(r_1)\psi_d(r_2)d\mathbf{r}_1d\mathbf{r}_2. \]
The first-order Coulomb correction is important because it excludes
double counting of the energy of the Coulomb interaction between electrons.
Indeed, the energy of the Coulomb interaction between electrons $i$ and $j$
is included in both $\epsilon_i$ and $\epsilon_j$.

\begin{table}
\caption{Relative change (in per cents) of the energy of the 1s state due to change
of nuclear radius by 10\%: (1) integration of
Eq.~(\ref{nuc}) with Coulomb 1s functions; (2) integration of Eq.~(\ref{nuc})
with the functions obtained by numerical solution of the Dirac equation with
finite size nucleus; (3) the difference between solutions of the Dirac
equation for two nuclear radii; (4) formula from Eq. (\ref{nu}).}
\label{tb:f}
\begin{ruledtabular}
\begin{tabular}{rcccc}
\multicolumn{1}{c}{$Z$} &
           (1)   &    (2)    &   (3)  &   (4)   \\
\hline
 80   &    0.01   &   0.01   &  0.01  &   0.01 \\
 90   &    0.03   &   0.02   &  0.02  &   0.02 \\
100   &    0.07   &   0.04   &  0.04  &   0.04 \\
110   &    0.16   &   0.08   &  0.08  &   0.09 \\
120   &    0.47   &   0.17   &  0.16  &   0.23 \\
130   &    1.74   &   0.35   &  0.32  &   0.82 \\
\end{tabular}
\end{ruledtabular}
\end{table}

The results for $Z$ from 80 to 160 are presented in Table~\ref{tb:e}.
In this table we present the change of the total electron energy of
the atom and corresponding change in the fission probability when 
nuclear radius changes by 10\%. Corrections to the fission probabilities
 were calculated at
$\hbar \omega=0.5$~MeV using formula (\ref{Pe}). At $\hbar \omega=1$~MeV
the change of the probability is about 2 times smaller as is obvious from
Eq.~(\ref{Pe}). Eqs. (\ref{P2}) and (\ref{Pe}) start to deviate significantly
 from
each other at about $Z=130$. In the vicinity of the {\em stability island}
($Z \approx 120$ to 130) electrons decrease fission probability by 
few per cent.
The effect is large at extremely high $Z$ leading to doubling of the nuclear
lifetime at $Z=160$. The $1s$ electrons give about 30\% of the total
energy and more than a half of its change due to the change of the nuclear
radius. This is important because the total energy depends on the
electron configuration and accurate calculations should include
investigation which configuration corresponds to the ground state.
Large contribution from the $1s$ electrons show that the effect
of external electrons can be neglected.

Since the change of the electron energy is dominated by 1s electrons it is 
instructive to consider a simple picture with only one electron in the
1s state. The change of its energy due to the change of the nuclear 
radius is very small. Therefore it is natural to try perturbation 
theory. First order correction to the Coulomb energy can be written as
\begin{equation}
  \Delta E(r_N) = -e\int(\phi - Ze/r)\psi^2(r)dV,
\label{nuc}
\end{equation}
where $\phi$ is the nuclear potential corresponding to the finite
nuclear radius $r_N$. The change of the energy due to small change
of nuclear radius can be found as the difference
\begin{equation}
  \delta (\Delta E) = \Delta E(r_{NB}) - \Delta E(r_{NA}).
\label{delta}
\end{equation}
Here $r_{NA}$ and $r_{NB}$ are nuclear radii at points A and B of Fig.~\ref{fig}.
The results for the relative change of the energies of 1s electrons
for different $Z$ obtained with the use of the relativistic Coulomb 
wave functions in the Eq.~(\ref{nuc}) are presented in second column
of Table~\ref{tb:f}. The change of nuclear radius is 10\% and the
radii are the same as in Table~\ref{tb:e}. The third column of Table~\ref{tb:f}
presents similar results which were obtained by replacing the Coulomb
wave functions with the numerical solutions of the Dirac equation 
corresponding to finite nuclear radius. The results are very close
for small $Z$ but deviate significantly for higher $Z$. This actually
means that the perturbation theory doesn't work in spite of the fact that
the correction to the energy is very small. The reason for this is
very simple: the perturbation is not small but it is localized in the
very small volume of space limited by the nuclear radius. Indeed, only
the nuclear volume contributes to the integral (\ref{nuc}). The finite
nuclear radius dramatically changes the electron wave function inside
this volume. 
In the end we use accurate numerical calculations rather than perturbation
theory to find the change of electron energy due to change of nuclear
radius.

In fourth column of Table~\ref{tb:f} we present the results of the 
numerical solution of the Dirac equation with the finite 
nuclear size. These results are in good agreement with the data
in previous column. The energy shift is dominated by the
 s-wave electrons.  Note that the single-electron s-wave energy shift due
  to finite nuclear size can be approximated by a semi-empirical 
formula which is accurate to few per cent from $Z=1$ up to
$Z=100$:
\begin{equation}
  \Delta E_{\nu} \approx \frac{|E_{\nu}|}{\nu}\frac{2}{5\gamma(\gamma+1)}
  \left(2Z\frac{r_N}{a_0}\right)^{2\gamma},
\label{nu}
\end{equation}
where $\nu$ is the effective principal quantum number 
($\nu=Z_{eff}\sqrt{a.u./(2|E_{\nu}|)}$;
$Z_{eff}=Z$, $\nu=1$ for the 1s state in the Coulomb field; $Z_{eff}=1$,
 $\nu \sim 1.7$ for
 an external electron in a neutral atom), $\gamma=\sqrt{1-(Z\alpha)^2}$,
$a_0$ is Bohr radius. If change of radius is small ($\delta r_N \ll r_N$) 
then Eq.~(\ref{nu}) leads to the following expression for the volume 
isotope shift
\begin{equation}
  \delta(\Delta E_{\nu}) \approx \frac{|E_{\nu}|}{\nu}\frac{4}{5(\gamma+1)}
  \left(2Z\frac{r_N}{a_0}\right)^{2\gamma}\frac{\Delta r_N}{r_N}.
\label{dnu}
\end{equation}
The energy shifts calculated with the use of Eqs.~(\ref{nu}) and (\ref{dnu})
are presented in last column of Table~\ref{tb:f}. They are in good agreement 
with the accurate numerical solutions of the previous column.

  To perform an accurate nuclear calculation of the fission probability
change due to the electron potential it is convenient to have an analytical
 formula for the electron energy as a function of the  nuclear radius.
Therefore, we present the result of the fitting of the calculated
total electron energy of a neutral atom as function of the nuclear
charge $Z$ and nuclear radius $r_N$:
\begin{eqnarray}
  & E_{total}=E_0\left[1-\frac{2}{5\gamma(\gamma+1)}\left(2Z\frac{r_N}{a_0}\right)^{2\gamma} 
\nonumber \right.\\
  & \times \left. \left[0.416-0.002(Z-100)\right]\right],
\label{etotal}
\end{eqnarray}
where $E_0$ does not depend on nuclear radius:
\begin{eqnarray}
  & E_0 = 1.266 Z^{\frac{1}{3}}(\gamma-1)mc^2\left[1-1.52\times 10^{-3}(Z-100) \right. \nonumber \\
  & \left. -2.8\times 10^{-5}(Z-100)^2\right].
\label{E0}
\end{eqnarray}
These formulas can be used in nuclear calculations of the fission process.
Note that here we assume Eq. (\ref{rho}) for the nuclear charge density
$\rho(r)$ averaged over nuclear rotation. This formula contains the fixed
parameter $D$ and the variable parameter $r_N$. Nuclear fission calculations
operate with different variable parameters (e.g. the deformation parameters).
 Relation between $r_N$
and these parameters may be easily  established by a numerical comparison
of the mean squared nuclear charge radii. It is assumed
that  nuclear physicists calculate $r_N$ for each intermediate state
of the nucleus during the fission process and then calculate nuclear response
to the electron potential for a given $r_N$. To avoid misunderstanding,
one does not need to calculate the fission lifetime to high accuracy
to make use of the electron effect. One only should calculate the
 (relative) change of the lifetime when the electron potential is added.    

 The effect of electrons on the alpha and proton emission
 barriers are  much smaller. However, one can use the same Eqs.(11),(12)
 to estimate them.

  Here we assume that the experiment consists in comparison of the fission
probabilities for bare nucleus and neutral atom. If necessary, we can provide
the result for an ion with an arbitrary number of electrons.

In conclusion we state that we have calculated the change of the total electron
energy of heavy and superheavy atoms due to a change of nuclear radius and
fitted the results with simple analytical formula. This formula can be 
used in nuclear calculations to include the effect of atomic electrons on the
probabilities of nuclear fission. Simple estimates based on the  parabolic
fission barrier show that although the effect is small it is probably
detectable in some nuclei. The small value of the effect has its advantage:
it is enough to consider a linear response of a nucleus  to  the probe -
the small $r_N$-dependent part of the potential produced by the atomic
electrons.

This opens a new way of studying the fission process.


\begin{thebibliography}{999}

\frenchspacing

\bibitem{Greiner} W. Greiner, Adv. Quant. Chem. {\bf 53}, 99 (2008).

\bibitem{Oganessian} Y. Oganessian, J. Phys. G {\bf 34}, R165 (2007).

\bibitem{Cwiok} S. Cwiok, P. H. Heenen, and W. Nazarewicz,
Nature {\bf 433}, 705 (2005).

\bibitem{Ackermann} D. Ackermann, Eur. Phys. J. A {\bf 25}, 577 (2005).


\bibitem{Leander} D.F. Zaretski and V.M. Novikov, Nucl. Phys.{\bf 28}, 177 (1961).
 C. Leander and P.Moller.Phys. Lett B {\bf 57}, 245 (1975).

\bibitem{Segre} E. Segr\'{e}, {\em Nuclei and Particles}, Second Edition,
The Benjamin/Cummings Publishing Company, Inc.,London (1977).

\bibitem{Bohr} A. Bohr and B. Mottelson, {\em Nuclear Structure},
W. A. Benjamin, Inc., New York (1974). 


\end{thebibliography}
\end{document}